\documentclass[floatfix,aps,amsmath,nofootinbib,twocolumn,10pt]{revtex4}

\usepackage{listings}
\usepackage{graphicx}
\usepackage{bm}
\usepackage{rotating}
\usepackage{array}
\usepackage{amsmath}
\usepackage{amssymb} %花体字母加粗
\usepackage{mathrsfs} %花体字母
\usepackage{cancel}
\usepackage{subfig}
\usepackage{float}
\usepackage{caption}

\def\({\left(}
\def\){\right)}
\def\[{\left[}
\def\]{\right]}

\def\e{\begin{equation}}
\def\q{\end{equation}}
\def\m{\begin{eqnarray}}
\def\n{\end{eqnarray}}

\begin{document}
\title{Forecasting constraints on the no-hair theorem from the stochastic gravitational wave background}% Force line breaks with \\
\author{Chen Tan$^{1,2,3}$}
\author{Ke Wang$^{1,2,3}$}
\thanks{Corresponding author: {wangkey@lzu.edu.cn}}
\affiliation{$^1$Institute of Theoretical Physics $\&$ Research Center of Gravitation, Lanzhou University, Lanzhou 730000, China}
\affiliation{$^2$Key Laboratory of Quantum Theory and Applications of MoE, Lanzhou University, Lanzhou 730000, China}
\affiliation{$^3$Lanzhou Center for Theoretical Physics $\&$ Key Laboratory of Theoretical Physics of Gansu Province, Lanzhou University, Lanzhou 730000, China}

\date{\today}% It is always \today, today,

\begin{abstract}
Although the constraints on general relativity (GR) from each individual gravitational-wave (GW) event can be combined to form a cumulative estimate of the deviations from GR, the ever-increasing number of GW events used also leads to the ever-increasing computational cost during the parameter estimation. Therefore, in this paper, we will introduce the deviations from GR into GWs from all events in advance and then create a modified stochastic gravitational-wave background (SGWB) to perform tests of GR. More precisely, we use the $\mathtt{pSEOBNRv4HM\_PA}$ model to include the model-independent hairs and calculate the corresponding SGWB with a given merger rate. Then we turn to the Fisher information matrix to forecast the constraints on the no-hair theorem from SGWB at frequency $10[{\rm Hz}]\lesssim f\lesssim10^3[{\rm Hz}]$ detected by the third-generation ground-based GW detectors, such as the Cosmic Explorer. We find that the forecasting constraints on hairs at $68\%$ confidence range are $\delta\omega_{220}=0\pm0.1296$ and $\delta\tau_{220}=0\pm0.0678$ when the flat priors about the merger rate are added but $\delta\omega_{220}=0\pm0.0903$ and $\delta\tau_{220}=0\pm0.0608$ when the non-flat priors about the merger rate are added.
\end{abstract} 

\maketitle

%\tableofcontents

\section{Introduction}
\label{sec:intro}
There are about 90 compact binary coalescences observed by the LIGO Scientific Collaboration~\cite{LIGO}, the Virgo Collaboration~\cite{Virgo} and the KAGRA Collaboration~\cite{KAGRA} during the first three observing runs~\cite{LIGOScientific:2018mvr,LIGOScientific:2020ibl,LIGOScientific:2021djp}. Through these observed gravitational-wave (GW) transient events, the full population of merging compact binaries surrounding us can be further inferred~\cite{KAGRA:2021duu,LIGOScientific:2020kqk}.
Consequently, a superposition of GWs from the surrounding population of these astrophysical sources will create a stochastic gravitational-wave background (SGWB) with energy density $\Omega_{\rm GW}(f)\sim 10^{-9}$ at frequency $10[{\rm Hz}]\lesssim f\lesssim10^3[{\rm Hz}]$~\cite{KAGRA:2021duu}. Due to the limited sensitivity of the current ground-based GW detectors, however, no evidence for a SGWB at this frequency span was found~\cite{KAGRA:2021kbb,LIGOScientific:2019vic,LIGOScientific:2016jlg}. Encouragingly, there are multiple lines of evidence for an excess SGWB signal with amplitude $A_{\rm SGWB}\sim 10^{-14}$ at frequency $f\sim10^{-8}[{\rm Hz}]$ recently in the NANOGrav 15-year data~\cite{NANOGrav:2023gor} and the second data release from EPTA~\cite{EPTA:2023fyk}. Therefore, there is an excellent probability that a SGWB at frequency $10[{\rm Hz}]\lesssim f\lesssim10^3[{\rm Hz}]$ can be detected by the third-generation ground-based GW detectors, such as the Cosmic Explorer (CE)~\cite{Reitze:2019iox} and the Einstein Telescope (ET)~\cite{Punturo:2010zz}.

The ever-increasing number of detections of GWs from compact binaries by the current ground-based GW detectors allows the ever-more sensitive tests of general relativity (GR) with GW generation, propagation or polarization~\cite{LIGOScientific:2021sio,LIGOScientific:2020tif,LIGOScientific:2019fpa}. More precisely, one can perform residuals test~\cite{LIGOScientific:2016lio,LIGOScientific:2020ufj}, inspiral–merger–ringdown (IMR) consistency test~\cite{Ghosh:2016qgn,Ghosh:2017gfp}, extra polarization contents searches~\cite{Pang:2020pfz,Takeda:2020tjj}, echoes searches~\cite{Conklin:2017lwb,Uchikata:2023zcu} as well as parametrized tests of GW generation and propagation~\cite{Yunes:2009ke,Nishizawa:2017nef,Meidam:2017dgf}, spin-induced multipole moment effects~\cite{Carter:1971zc,Gurlebeck:2015xpa}, modified GW dispersion relation ~\cite{Will:1997bb,Mirshekari:2011yq,Wang:2020pgu,Zhao:2022pun,Niu:2022yhr,Wang:2021ctl} and the no-hair theorem~\cite{Isi:2019aib,Carullo:2019flw,Wang:2021elt}. All of there tests will introduce some ad hoc parameters to account for the deviations from GR. Since these new-introduced parameters are independent of the individual sources by construction and cleanly encode information about the underlying theory of gravity, the constraints on them from each individual GW event can be combined to form a cumulative estimate of the deviations from GR. The more GW events are taken into consideration, the tighter cumulative constraints are obtained.
However, the ever-increasing number of GW events used also leads to the ever-increasing computational cost during the parameter estimation.

To avoid the time-consuming parameter estimation procedure, one can combine GWs from all events in advance and then perform tests of GR. That is to say, we can use SGWB at frequency $10[{\rm Hz}]\lesssim f\lesssim10^3[{\rm Hz}]$ to test GR. While the information of GW propagation may wash out in SGWB, the information of GW generation and polarization must survive SGWB and some deviations from GR's GW generation and polarization will lead to some suppressions or enhancements in SGWB.
If such suppressions or enhancements have no particular distinguishing features, testing GR with SGWB must deals with the degeneracy between models. However, if one would not confine oneself to specific modified gravities, one can search for extra polarization contents in SGWB model-independently~\cite{LIGOScientific:2018czr,Jiang:2022uxp}. Similarly, in this paper, we will probe the deviations from GR's GW generation with SGWB directly. More precisely,
we will take the violation of the no-hair theorem as a heuristic example. 

\begin{figure*}[]
\begin{center}
\subfloat{\includegraphics[width=0.8\textwidth]{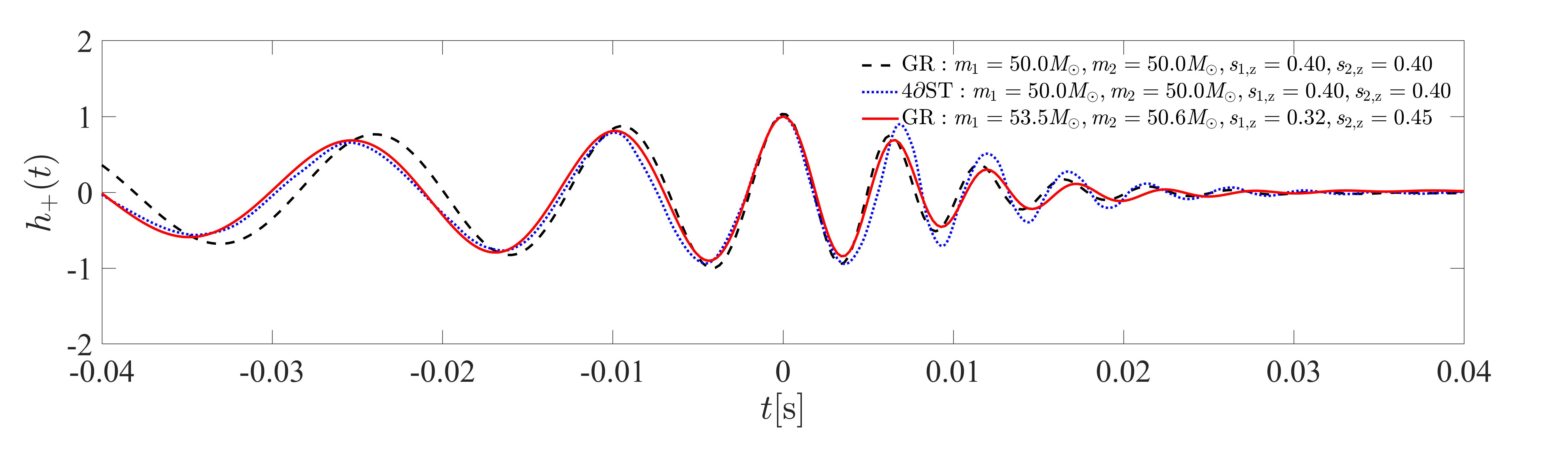}}
\end{center}
\captionsetup{justification=raggedright}
\caption{The IMR time-domain waveforms predicted by GR and four-derivative scalar-tensor theories ($4\partial \rm{ST}$) ~\cite{AresteSalo:2023mmd}. The black dashed curve is plotted with $\{m_{1}=m_{2}=50M_{\odot},s_{1,\rm{z}}=s_{2,\rm{z}}=0.40\}$ in GR and its $4\partial \rm{ST}$ counterpart (the blue dotted curve) is also given. The effects of the extra scalar hairs from $4\partial \rm{ST}$ on the inspiral and merger regions can be summarized by a set of effective intrinsic parameters $\{m_{1}=53.5M_{\odot},m_{2}=50.6M_{\odot},s_{1,\rm{z}}=0.32,s_{2,\rm{z}}=0.45\}$ in GR and the residual effects of these extra scalar hairs just appear in ringdown region (the red solid curve). More precisely, there are obvious differences between the red solid curve and the blue dotted curve in ringdown region while there is an almost complete overlap between these two curves before merger.}
\label{fig:assump}
\end{figure*}

Under the no-hair theorem, the ringdown frequencies and damping times of a perturbed Kerr black hole (BH) in GR can be predicted by its mass and dimensionless spin~\cite{Berti:2005ys,Isi:2021iql}.
Without the no-hair conjecture, a perturbed Kerr BH's ringdown frequencies and damping times are also dependent on the extra hairs/parameters~\cite{Isi:2019aib,Carullo:2019flw,Wang:2021elt}.
In fact, violating the no-hair theorem may also modify the whole IMR waveform for a BH binary. Such overall modifications will degenerate with other overall effects, for example changing the total mass of the BH binary, hence the final mass of its remnant Kerr BH.
That is to say, when the extra hairs dominate the inspiral and merger regions of a BH binary, the mass and dimensionless spin of its remnant Kerr BH can't be predicted by its intrinsic parameters only, hence an unknown ringdown region.
Therefore, for simplicity, we assume that the effects of the extra hairs on the inspiral and merger regions can be summarized by a set of effective intrinsic parameters under the no-hair theorem and then the residual effects of these extra hairs will just appear during an effective ringdown region. 
It is generally the case when different modified gravity theories are considered.
Here we validate our above assumption with a specific modified theory of gravity, as shown in Fig.~\ref{fig:assump}.
Finally the effective complete IMR time-domain waveform can be described by $\mathtt{pSEOBNRv4HM\_PA}$ model~\cite{Ghosh:2021mrv,Mihaylov:2021bpf} explicitly. 

This paper is organized as follows.
In section~\ref{sec:Ogw}, we calculate the stochastic gravitational wave background at frequency $10[{\rm Hz}]\lesssim f\lesssim10^3[{\rm Hz}]$ without the no-hair conjecture.
In section~\ref{sec:Con}, we forecast the constraints on the no-hair theorem from SGWB.
Finally, a brief summary and discussions are included in section~\ref{sec:SD}.

\section{Stochastic gravitational wave background without the no-hair conjecture}
\label{sec:Ogw}
\subsection{Analytical  Calculation}
\label{ssec:an}
The radiative degrees of freedom in transverse-traceless (TT) gauge can be written
in terms of two polarizations $h_{+}$ and $h_{\times}$:
\begin{equation}\label{eqn-1} 
%&h_{ij}=h_{+}(e_{+})_{ij}+h_{\times}(e_{\times})_{ij},
\begin{aligned}
h_{ij}=\begin{pmatrix}
h_{+} & h_{\times} &0 \\
h_{\times} & -h_{+} &0\\
0 & 0 &0
\end{pmatrix}
.
\end{aligned}
\end{equation}
After Fourier transform, their representation in the frequency domain is $\tilde{h}_{ij}=\mathcal{F}(h_{ij})$
\begin{equation}
% \tilde{h}_{ij}=\tilde{h}_{+}(e_{+})_{ij}+\tilde{h}_{\times}(e_{\times})_{ij},
\tilde{h}_{ij}=\begin{pmatrix}
\tilde{h}_{+} & \tilde{h}_{\times} &0 \\
\tilde{h}_{\times} & -\tilde{h}_{+} &0\\
0 & 0 &0
\end{pmatrix}
.
\end{equation}
Then the energy-momentum tensor of GWs is
\begin{equation} 
T_{\mu\nu}=\frac{c^{4}}{32\pi G} \left \langle \partial_{\mu}h_{ij} \partial_{\nu}h^{ij}\right \rangle,
\end{equation}
where $\left\langle ... \right \rangle$ indicates averaging over several wavelengths or periods. The energy density of GWs is just the $00$ component. So the energy flux through a sphere of radius $R$ is
\begin{equation} 
\frac{dE}{dt}=\frac{c^{3}R^{2}}{32\pi G} \int d\Omega  \left \langle \dot{h}_{ij} \dot{h}^{ij} \right \rangle, 
\end{equation}
and the total energy emitted is
\begin{equation} 
E=\frac{c^{3}R^{2}}{32\pi G}\int d\Omega \int_{-\infty}^{\infty}dt\ 
\dot{h}_{ij}\dot{h}^{ij},
\end{equation}
where dots denote derivative with respect to physical time $t$ and $d\Omega$ is a solid angle element.
Here we are considering all time. So $\left\langle ... \right \rangle$ disappears in the integral with respect to $t$.
We can rewrite the total energy in the frequency domain as
\begin{equation}
E=\frac{\pi c^{3}R^{2}}{4G}\int d\Omega \int _{0}^{\infty} df   f^{2} \tilde{h}^{ij}(f)\tilde{h}^{*}_{ij}(f).
\end{equation}
So the energy spectrum is given by
\begin{eqnarray} 
\nonumber
\frac{dE}{df}&=&\frac{\pi c^{3}R^{2}f^{2}}{4G}\int d\Omega \tilde{h}^{ij}(f)\tilde{h}^{*}_{ij}(f)\\
&=&\frac{\pi c^{3}R^{2}f^{2}}{2G}\int d\Omega  \left ( \left |\tilde{h}_{+}(f)\right |^{2}+\left |\tilde{h}_{\times}(f)\right |^{2}\right ).
\end{eqnarray}

Next we will do the integral with respect to $d\Omega$. The general GW signal can be described as a linear combination of the two polarization states or expressed in terms of series of spin-weighted spherical harmonics
\begin{eqnarray}
\label{eq:h}
\nonumber
\textbf{h}&\equiv& h_{+}-ih_{\times}\\
 &=&\sum_{lm}\textbf{h}_{lm}Y^{-2}_{lm}(\varphi,\iota),
\end{eqnarray}
where $\varphi$ is the azimuthal direction to the observer and $\iota$ is the inclination angle.
Here we use the spin-weighted spherical harmonics to describe the whole IMR waveform including the ringdown waveform even thought the spin-weighted spheroidal harmonics are
more accurate for the remnant Kerr BHs. In fact, the spherical version is a sufficiently good approximation of its spheroidal counterpart~\cite{Isi:2021iql,Berti:2005gp}.
For the two dominant modes with $l=2$ and $m=\pm2$, the spin-weighted spherical harmonics are~\cite{Ajith:2007kx,Wiaux:2005fm}
\begin{eqnarray}
\nonumber
Y^{-2}_{2-2}(\varphi,\iota)&\equiv \sqrt{\frac{5}{64\pi}}(1-\cos\iota)^{2}e^{-2i\varphi},\\
Y^{-2}_{22}(\varphi,\iota)&\equiv \sqrt{\frac{5}{64\pi}}(1+\cos\iota)^{2}e^{2i\varphi}.
\end{eqnarray}
When $\varphi=0$ and $\iota=0$, for example, we have the GW signal as
\begin{equation}
\label{eq:h2}
\textbf{h}(\varphi=\iota=0)\approx 4\sqrt{\frac{5}{64\pi}}\textbf{h}_{22}.
\end{equation}
According to Eq.~(\ref{eq:h}), we rewrite 
\begin{equation} 
\frac{dE}{df}\approx \frac{\pi c^{3}R^{2}f^{2}}{2G}\int d\Omega\left( \left | \tilde{\textbf{h}}_{22}Y^{-2}_{22}(\varphi,\iota)\right |^{2} +\left |  \tilde{\textbf{h}}_{2-2}Y^{-2}_{2-2}(\varphi,\iota)\right |^{2} \right).
\end{equation}
Due to the orthonormality of spin-weighted spherical harmonics, the integral can be done easily
\begin{eqnarray} 
\nonumber
\frac{dE}{df}&\approx&\frac{\pi c^{3}R^{2}f^{2}}{2G}\left ( \left|\tilde{\textbf{h}}_{22}\right|^{2}+\left|\tilde{\textbf{h}}_{2-2}\right|^{2}\right )\\
&\approx&\frac{\pi c^{3}R^{2}f^{2}}{G}\left| \tilde{\textbf{h}}_{22}\right|^{2},
\end{eqnarray}
where we have used the equatorial symmetry $\textbf{h}_{22}=(\textbf{h}_{2-2})^{*}$.
According to Eq.~(\ref{eq:h}) and Eq.~(\ref{eq:h2}), we can express the energy spectrum approximately as
\begin{eqnarray}
\label{eq:dedfend} 
\frac{dE}{df}&\approx&\frac{4\pi^{2} c^{3}R^{2}f^{2}}{5G}\left| \tilde{\textbf{h}}(\varphi=\iota=0)\right|^2\\
\nonumber
 &\approx&\frac{4\pi^{2} c^{3}R^{2}f^{2}}{5G}\left (\left| \tilde{h}_{+}(\varphi=\iota =0)\right|^2+\left|\tilde{h}_{\times}(\varphi=\iota=0)\right|^2\right ).
\end{eqnarray}
Although here we have fixed the two angles $\varphi=\iota=0$, the values of them don't affect the total energy emitted.
When the higher-order modes are taken into consideration, we can express the energy spectrum more accurately as
\begin{eqnarray} 
\label{eq:hm}
\nonumber
\frac{dE}{df}&=&\frac{\pi c^{3}R^{2}f^{2}}{2G}\sum_{lm} \int d\Omega \left | \tilde{\textbf{h}}_{lm}Y^{-2}_{lm}(\varphi,\iota)\right |^{2} \\
&=&\frac{\pi c^{3}R^{2}f^{2}}{G}\sum_{l|m|}\left| \tilde{\textbf{h}}_{lm}\right|^{2},
%\tilde{\textbf{h}}_{lm}&=&\frac{\tilde{\textbf{h}}_{lm}Y^{-2}_{lm}(\varphi,\iota)}{Y^{-2}_{lm}(\varphi,\iota)} ,
\end{eqnarray}
where we have used the equatorial symmetry of all modes $(l,|m|)$ and $\tilde{\textbf{h}}_{lm}$ can be numerically obtained because $\tilde{\textbf{h}}_{lm}Y^{-2}_{lm}(\varphi,\iota)$ as a whole can be numerically obtained at fixed angles, for example $\varphi=0$ and $\iota=\pi/2$.
%$\tilde{\textbf{h}}_{lm}Y^{-2}_{lm}(\varphi,\iota)$ are the contributions of every modes to the whole waveform $\tilde{\textbf{h}}$ which are given numerically at fixed angles $\varphi=0$ and $\iota=\pi/2$. 
Unless otherwise stated, in our following calculations, we used the $(2,2)$ mode only. 

The dimensionless GW energy density per logarithmic frequency interval is 
\begin{equation} 
\Omega_{\rm{GW}}=\frac{f}{\rho_{c}}\frac{d\rho_{\rm{GW}}}{df},
\end{equation}
where $\rho_{c}=3H_{0}^{2}c^{2}/8\pi G$ is the critical energy density and $H_{0}=67.36[\rm{km\ s^{-1}Mpc^{-1}}]$~\cite{Planck:2018vyg}.
The contribution of BH binary mergers can be estimated as~\cite{KAGRA:2021kbb}
\begin{equation}
\label{eq:o_k} 
\Omega_{\rm{GW,BH}}=\frac{f}{\rho_{c}}\int_{0}^{z_{\rm{max}}}dz\frac{R_{\rm{BH}}(z)\left \langle dE_{s}/df_{s}\right \rangle_{\rm{BH}}}{(1+z)H(z)},
\end{equation}
where $f_{s}=f(1+z)$ is frequency in the source frame.
The Hubble parameter is $H(z)=H_{0}\sqrt{\Omega_{m}(1+z)^{3}+\Omega_{\Lambda}}$, where $\Omega_{m}=0.3153$ is the matter density parameter and $\Omega_{\Lambda}=0.6847$ is the dark energy density parameter~\cite{Planck:2018vyg}.   
The quantity $\left \langle dE_{s}/df_{s}\right \rangle_{\rm{GW,BBHs}} $
is the source-frame energy radiated by a single source, which should be averaged over the ensemble properties of the full BH binary population
\begin{equation}
\label{eq:mdf}
\left \langle dE_{s}/df_{s}\right \rangle_{\rm{BH}}  =\int d \alpha \ p_{\rm{BH}}(\alpha ) \frac{dE_{s}}{df_{s}}(\alpha),
\end{equation}
where $p_{\rm{BH}}(\alpha)$ is the probability distribution of intrinsic parameters $\alpha$ (e.g. masses, spins, etc.) for the full BH binary population.
The merger rate $R_{\rm{BH}}(z)$ can be obtained by a convolution of the BH binary formation rate $R_{\rm{BH}}^{f}(z_f)$ with the distribution of the time delays $P(t_d)$ between BH binary formation and merger~\cite{LIGOScientific:2016fpe,Cholis:2016xvo}.
Also $R_{\rm{BH}}(z)$ can be obtained from a merger rate $R_{\rm{BH}}(z_{\rm{quoted}})$ at a quoted redshift $z_{\rm{quoted}}=0$~\cite{Callister:2016ewt} or $z_{\rm{quoted}}=0.2$~\cite{KAGRA:2021duu}. 
Here we turn to the fiducial Power Law $+$ Peak (PP) model~\cite{LIGOScientific:2020kqk,KAGRA:2021duu},
as shown in Fig.~10 of~\cite{KAGRA:2021duu}, where $R_{\rm{BH}}(z_{\rm{quoted}}=0.2)=28.3[\rm{Gpc^{-3}yr^{-1}}]$ is obtained by integrating the primary mass distribution $\frac{\mathrm{d}R_{\rm{BH}}}{\mathrm{d}m_{1}}$ or the mass ratio distribution $\frac{\mathrm{d}R_{\rm{BH}}}{\mathrm{d}q}$. After integrating over the contribution of spins, $p_{\rm{BH}}(\alpha)\approx \frac{1}{R_{\rm{BH}}^2}\frac{\mathrm{d}R_{\rm{BH}}}{\mathrm{d}m_{1}}\frac{\mathrm{d}R_{\rm{BH}}}{\mathrm{d}q}$ serves as a good approximation.

\subsection{Numerical Calculation}
\label{ssec:nu}
Given the waveform of every GW event, we can calculate the total SGWB energy-density spectrum numerically according to Eq.~(\ref{eq:o_k}). Here we will consider the deviations from GR’s GW generation, namely violating the no-hair theorem.
First we turn to the $\mathtt{PyCBC}$ package\cite{Biwer:2018osg} and the $\mathtt{pSEOBNRv4HM\_PA}$ model~\cite{Ghosh:2021mrv,Mihaylov:2021bpf} to generate the IMR time-domain waveforms for every possible BH binary with different primary mass and mass
ratio allowed by a given fiducial PP model~\cite{LIGOScientific:2020kqk,KAGRA:2021duu} individually.
In the $\mathtt{pSEOBNRv4HM\_PA}$ model, the ad hoc hairs $\delta\omega_{lmn}$ and $\delta\tau_{lmn}$ are introduced as 
\begin{equation}
\label{eq:eqn-17} 
\begin{aligned}
\omega_{lmn}&=\omega_{lmn}^{\rm{GR}}(1+\delta \omega_{lmn}),\\
\tau_{lmn}&=\tau_{lmn}^{\rm{GR}}(1+\delta \tau_{lmn}),\\
\end{aligned}
\end{equation}
where $\omega_{lmn}$ and $\tau_{lmn}$ are the ringdown frequencies and damping times respectively.

Why we can fix the fiducial PP model for different values of $\{\delta \omega_{lmn}, \delta \tau_{lmn}\}$? The fiducial PP model, in our paper, just provides a effective mass distribution when we summarize the effects of $\{\delta \omega_{lmn}, \delta \tau_{lmn}\}$ on the inspiral and merger regions with a set of effective intrinsic parameters.
Given a modified theory of gravity, the effective mass distribution will vary with the values of $\{\delta \omega_{lmn}, \delta \tau_{lmn}\}$ to compensate for the changes of the inspiral and merger regions. Similarly, given the values of $\{\delta \omega_{lmn}, \delta \tau_{lmn}\}$, the effective mass distribution will also vary with the different modified theory of gravity to compensate for the corresponding changes of the inspiral and merger regions.
Under different situations of the extra hairs, however, the effective mass distribution provided by the fiducial PP model can be unchanged in certain cases. 
For example, we can fix the effective mass distribution as the realistic one picked out by the observations of GW transients during our calculations for different values of $\{\delta \omega_{lmn}, \delta \tau_{lmn}\}$ through adjusting the modified theory of gravity simultaneously. That is to say, our model-independent parameterization of Eq.~(\ref{eq:eqn-17}) is not confined to a specific modified theory of gravity, but aims to identify any deviation from GR.

\begin{figure*}[]
\begin{center}
\subfloat{\includegraphics[width=0.8\textwidth]{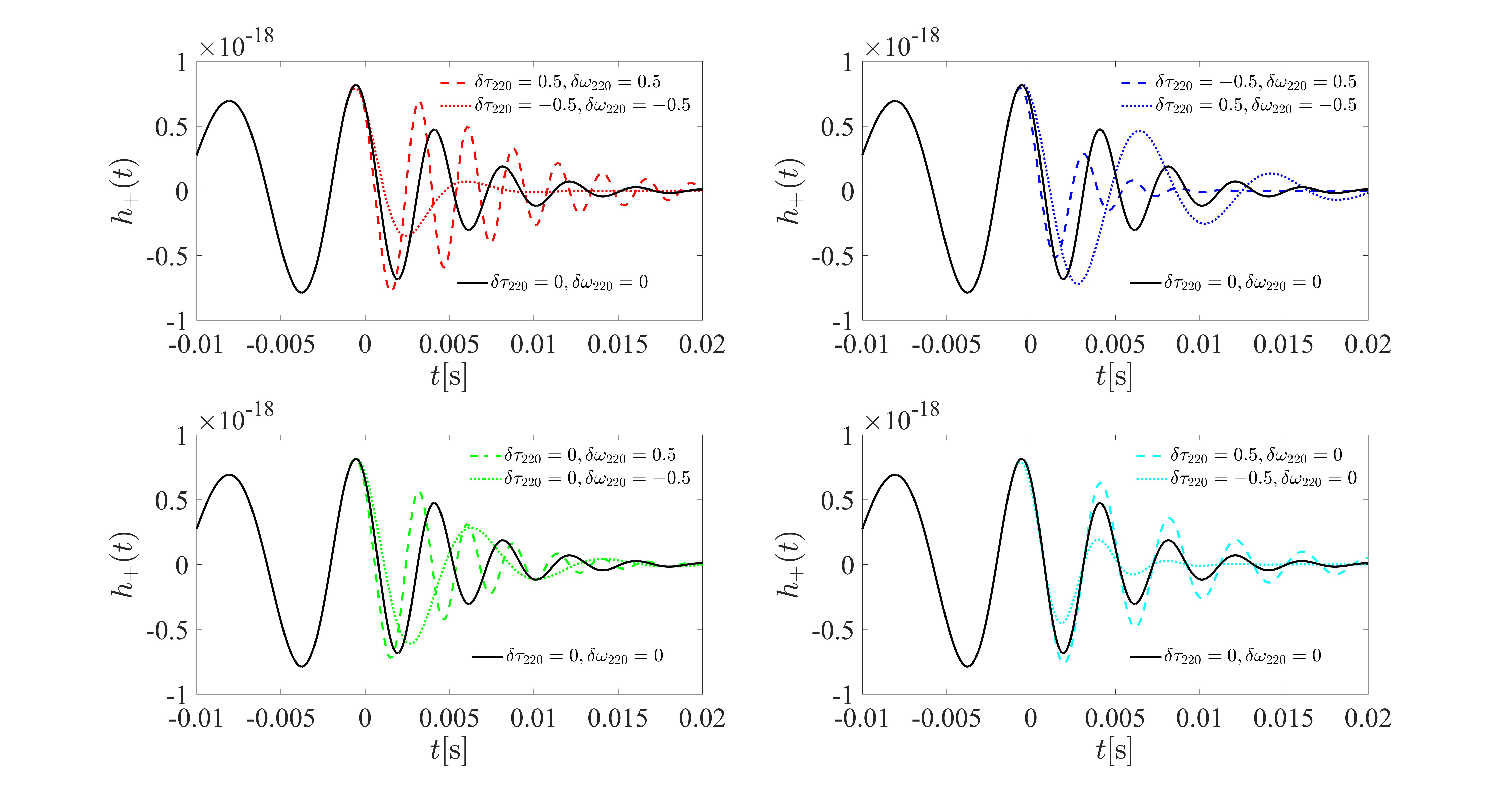}}
\end{center}
\captionsetup{justification=raggedright}
\caption{The IMR time-domain waveforms from a GW150914-like event for `+'–polarization given by the $\mathtt{pSEOBNRv4HM\_PA}$ model~\cite{Ghosh:2021mrv,Mihaylov:2021bpf}, where the luminosity distance to the binary is set as $1{\rm Mpc}$. The GR version with $\{\delta \tau_{220}=0,\delta \omega_{220}=0\}$ is the black solid curve; 
the non-GR version with $\{\delta \tau_{220}=0.5,\delta \omega_{220}=0.5\}$, $\{\delta \tau_{220}=-0.5,\delta \omega_{220}=-0.5\}$, $\{\delta \tau_{220}=-0.5,\delta \omega_{220}=0.5\}$, $\{\delta \tau_{220}=0.5,\delta \omega_{220}=-0.5\}$, $\{\delta \tau_{220}=0,\delta \omega_{220}=0.5\}$, $\{\delta \tau_{220}=0,\delta \omega_{220}=-0.5\}$, $\{\delta \tau_{220}=0.5,\delta \omega_{220}=0\}$ or $\{\delta \tau_{220}=-0.5,\delta \omega_{220}=0\}$ is plotted by the red dashed, red dotted, blue dashed, blue dotted, green dashed, green dotted, cyan dashed or cyan dotted curve respectively.}
\label{fig:gw150914_td}
\end{figure*}

\begin{figure*}[]
\begin{center}
\subfloat{\includegraphics[width=0.8\textwidth]{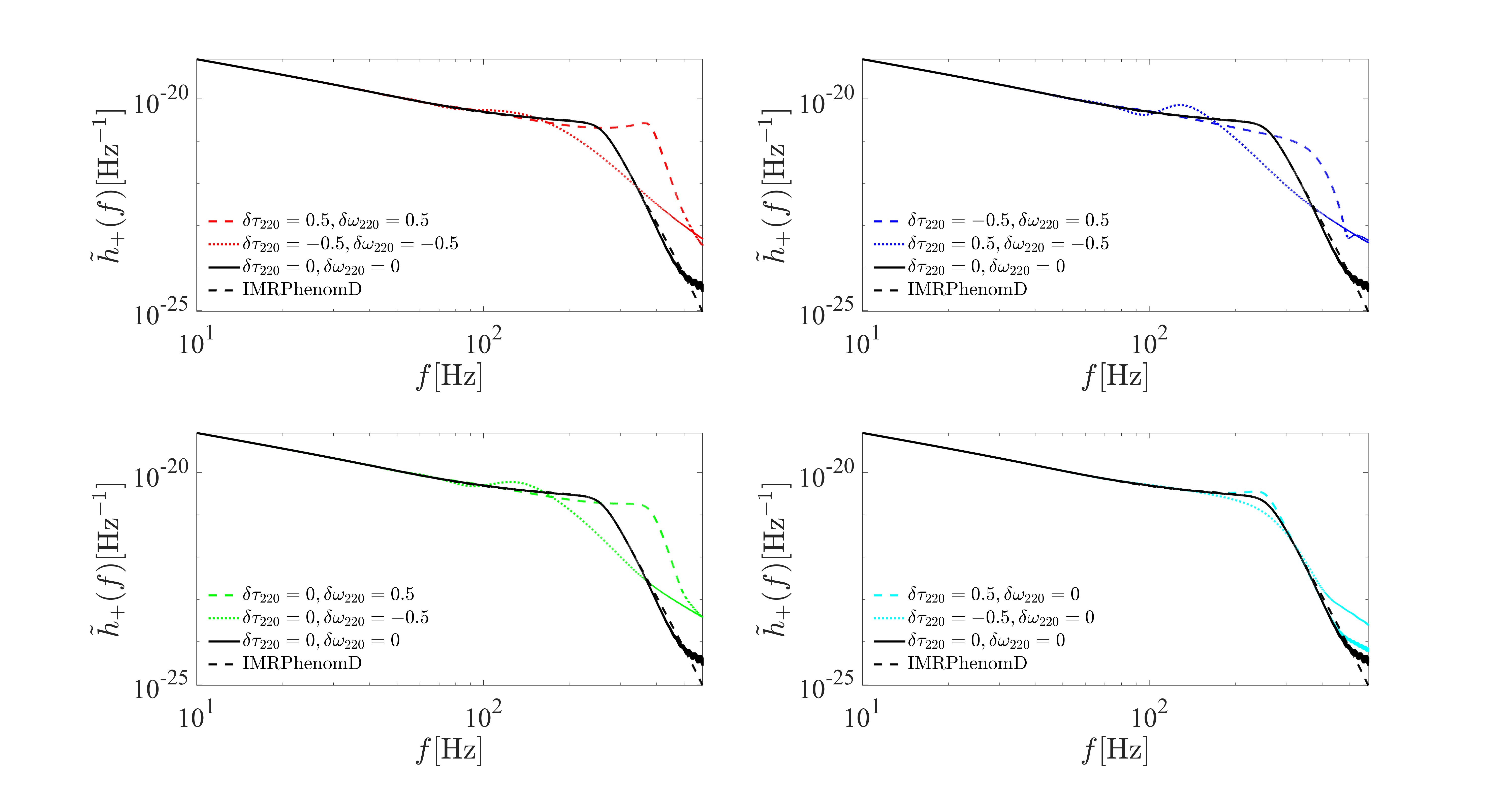}}
\end{center}
\captionsetup{justification=raggedright}
\caption{The IMR frequency-domain waveforms from a GW150914-like event for `+'–polarization calculated by the Fourier transform of Fig.~\ref{fig:gw150914_td}. The GR version with $\{\delta \tau_{220}=0,\delta \omega_{220}=0\}$ is the black solid curve; 
the non-GR version with $\{\delta \tau_{220}=0.5,\delta \omega_{220}=0.5\}$, $\{\delta \tau_{220}=-0.5,\delta \omega_{220}=-0.5\}$, $\{\delta \tau_{220}=-0.5,\delta \omega_{220}=0.5\}$, $\{\delta \tau_{220}=0.5,\delta \omega_{220}=-0.5\}$, $\{\delta \tau_{220}=0,\delta \omega_{220}=0.5\}$, $\{\delta \tau_{220}=0,\delta \omega_{220}=-0.5\}$, $\{\delta \tau_{220}=0.5,\delta \omega_{220}=0\}$ or $\{\delta \tau_{220}=-0.5,\delta \omega_{220}=0\}$ is plotted by the red dashed, red dotted, blue dashed, blue dotted, green dashed, green dotted, cyan dashed or cyan dotted curve respectively; the black dashed curve is the GR IMR frequency-domain waveform from a GW150914-like event for `+'–polarization and plotted directly by the $\mathtt{IMRPhenomD}$ model~\cite{Khan:2015jqa}.}
\label{fig:gw150914_fd}
\end{figure*}

\begin{figure*}[]
\begin{center}
\subfloat{\includegraphics[width=0.8\textwidth]{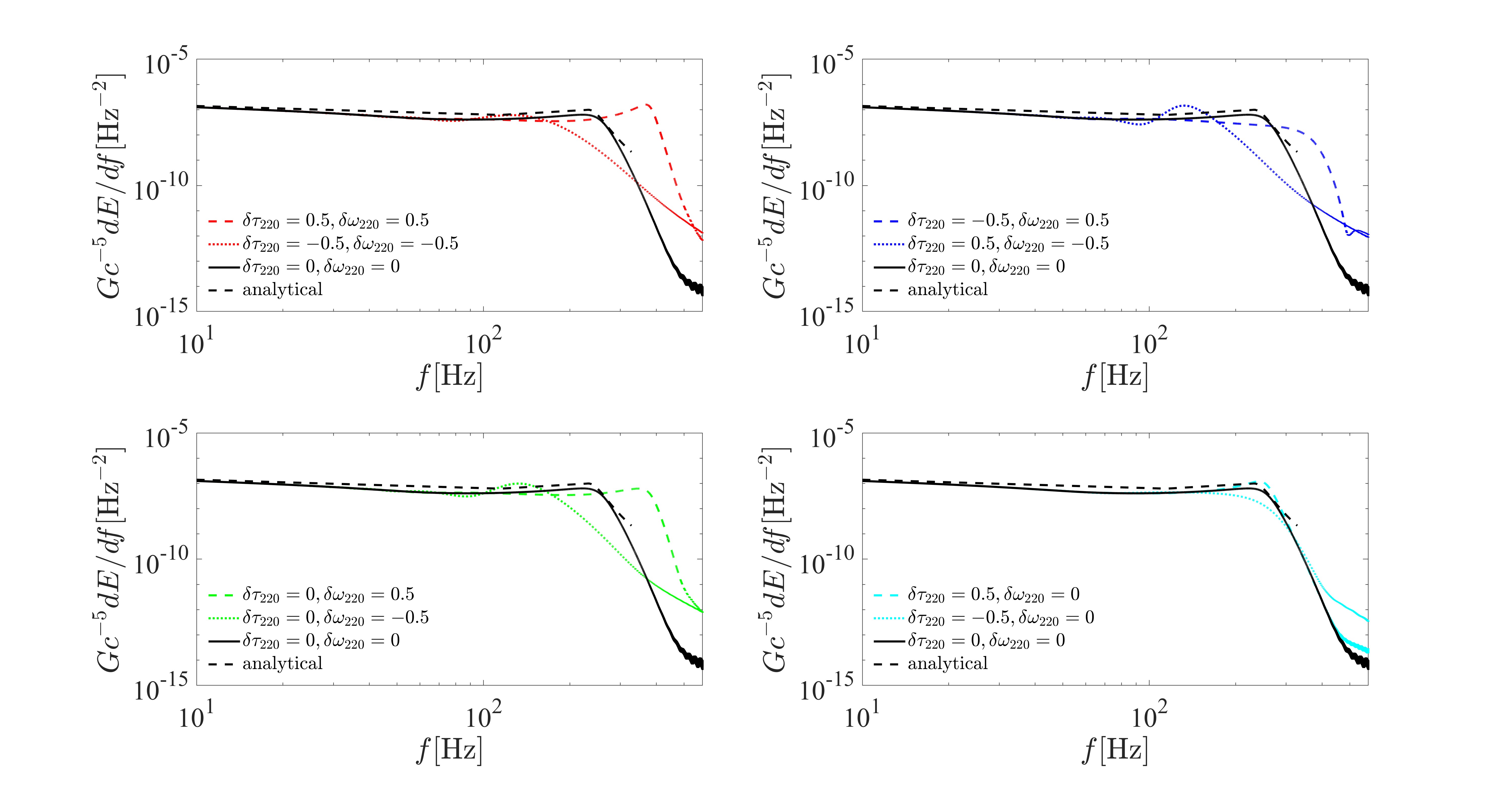}}
\end{center}
\captionsetup{justification=raggedright}
\caption{The energy spectrum for a GW150914-like event calculated numerically with the $\mathtt{pSEOBNRv4HM\_PA}$ model's~\cite{Ghosh:2021mrv,Mihaylov:2021bpf} waveforms. The GR version with $\{\delta \tau_{220}=0,\delta \omega_{220}=0\}$ is the black solid curve; 
the non-GR version with $\{\delta \tau_{220}=0.5,\delta \omega_{220}=0.5\}$, $\{\delta \tau_{220}=-0.5,\delta \omega_{220}=-0.5\}$, $\{\delta \tau_{220}=-0.5,\delta \omega_{220}=0.5\}$, $\{\delta \tau_{220}=0.5,\delta \omega_{220}=-0.5\}$, $\{\delta \tau_{220}=0,\delta \omega_{220}=0.5\}$, $\{\delta \tau_{220}=0,\delta \omega_{220}=-0.5\}$, $\{\delta \tau_{220}=0.5,\delta \omega_{220}=0\}$ or $\{\delta \tau_{220}=-0.5,\delta \omega_{220}=0\}$ is plotted by the red dashed, red dotted, blue dashed, blue dotted, green dashed, green dotted, cyan dashed or cyan dotted curve respectively; the black dashed curve is the analytical energy spectrum~\cite{Callister:2016ewt,Ajith:2007kx}.}
\label{fig:dedf150914_fd}
\end{figure*}

\begin{figure*}[]
\begin{center}
\subfloat{\includegraphics[width=0.8\textwidth]{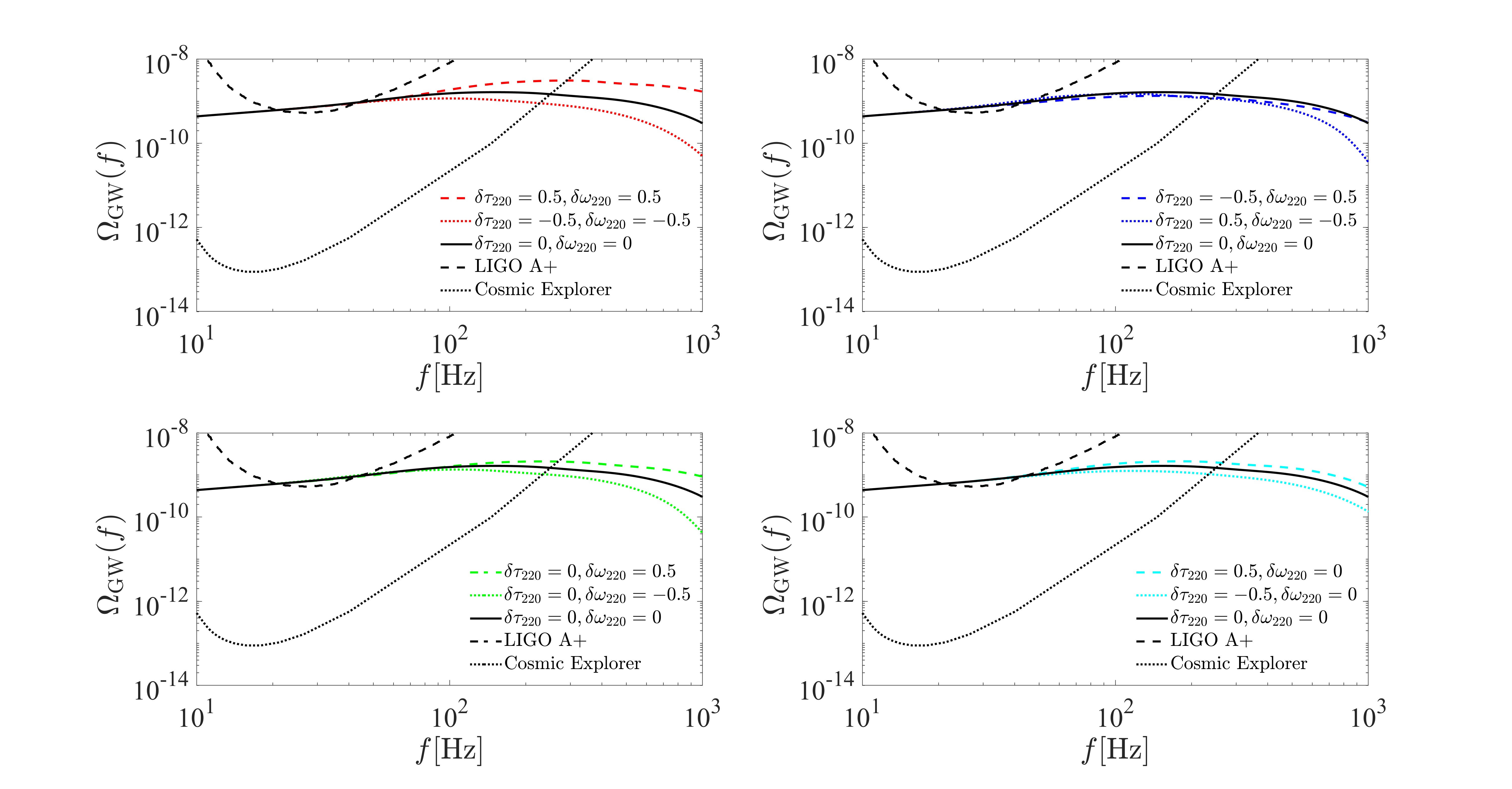}}
\end{center}
\captionsetup{justification=raggedright}
\caption{The total SGWB energy-density spectra at frequency $10[{\rm Hz}]\lesssim f\lesssim10^3[{\rm Hz}]$, where the ratio of the contribution of BH binary mergers to the contribution of NS binary and NS-BH mergers is about $1:0.3$~\cite{KAGRA:2021duu}. The GR version with $\{\delta \tau_{220}=0,\delta \omega_{220}=0\}$ is the black solid curve; 
the non-GR version with $\{\delta \tau_{220}=0.5,\delta \omega_{220}=0.5\}$, $\{\delta \tau_{220}=-0.5,\delta \omega_{220}=-0.5\}$, $\{\delta \tau_{220}=-0.5,\delta \omega_{220}=0.5\}$, $\{\delta \tau_{220}=0.5,\delta \omega_{220}=-0.5\}$, $\{\delta \tau_{220}=0,\delta \omega_{220}=0.5\}$, $\{\delta \tau_{220}=0,\delta \omega_{220}=-0.5\}$, $\{\delta \tau_{220}=0.5,\delta \omega_{220}=0\}$ or $\{\delta \tau_{220}=-0.5,\delta \omega_{220}=0\}$ is plotted by the red dashed, red dotted, blue dashed, blue dotted, green dashed, green dotted, cyan dashed or cyan dotted curve respectively; the black dashed curve is the expected sensitivity of LIGO’s anticipated ``A+'' configuration~\cite{Thrane:2013oya,Ogw:code,Curve:A+}; the black dotted curve is the sensitivity of CE~\cite{Curve:ce} by assuming one year's worth of data, a frequency bin width of $0.25[{\rm Hz}]$ and the same overlap reduction function as LIGO's.}
\label{fig:OBBHs_fd}
\end{figure*}

We plot the IMR time-domain waveforms from a GW150914-like event for ‘+’–polarization in Fig.~\ref{fig:gw150914_td}, where the luminosity distance is $1[{\rm Mpc}]$. 
By definition, $\delta\omega_{lmn}$ affects the ringdown frequencies (green curves) and $\delta\tau_{lmn}$ affects the ringdown damping times (cyan curves).
Then we make the Fourier transform of the IMR time-domain waveforms to get the IMR frequency-domain waveforms. We plot the IMR frequency-domain waveforms from a GW150914-like event for ‘+’–polarization in Fig.~\ref{fig:gw150914_fd}, where the black dashed curve is its GR counterpart and plotted directly by the $\mathtt{IMRPhenomD}$ model~\cite{Khan:2015jqa}. We find that the effects of $\delta\omega_{lmn}$ on the IMR frequency-domain waveform is lager than that of $\delta\tau_{lmn}$. Next, according to Eq.~(\ref{eq:dedfend}), we can get the energy spectrum for every BH binary. We plot the energy spectrum for a GW150914-like event in Fig.~\ref{fig:dedf150914_fd}, where the black dashed curve is the analytical energy spectrum~\cite{Callister:2016ewt,Ajith:2007kx}. Again we find that the effects of $\delta\omega_{lmn}$ on the energy spectrum is lager than that of $\delta\tau_{lmn}$. Finally, we plot the total SGWB energy-density spectrua at frequency $10[{\rm Hz}]\lesssim f\lesssim10^3[{\rm Hz}]$ in Fig.~\ref{fig:OBBHs_fd}, where the ratio of the contribution of BH binary mergers to the contribution of neutron star (NS) binary and neutron star–black hole (NS-BH) mergers is supposed to be $1:0.3$ which is consistent with the forecast by~\cite{KAGRA:2021duu}. Compared with Fig.~\ref{fig:gw150914_fd} and Fig.~\ref{fig:dedf150914_fd}, we confirm that the effects of $\delta\omega_{lmn}$ on the total SGWB energy-density spectrum is lager than that of $\delta\tau_{lmn}$. However, the frequency where their effects on the total SGWB energy-density spectrum become obvious is dependent on the values of them because SGWB results from the superposition of many GWs. For example, the blue dashed curve just deviates the black solid curve obviously at $f>10^3[{\rm Hz}]$.

It is worth noting that the $\mathtt{pSEOBNRv4HM\_PA}$ model is for spin-aligned compact binaries moving in quasi-circular orbits. For a given fiducial PP model, this approximation of the inspiral region will only affects the total SGWB energy-density spectra at the lower frequency span $10[{\rm Hz}]\lesssim f\lesssim10^2[{\rm Hz}]$ while the effects of $\{\delta \omega_{lmn}, \delta \tau_{lmn}\}$ dominate at the higher frequency span $10^2[{\rm Hz}]\lesssim f\lesssim10^3[{\rm Hz}]$. As shown in Fig.~5, these two parts don't correlate with each other directly and our forecasting constraints on $\{\delta \omega_{lmn}, \delta \tau_{lmn}\}$ will not be influenced by this approximation.

\section{Forecasting constraints on the no-hair theorem }
\label{sec:Con}
A stationary, Gaussian, unpolarized and isotropic SGWB can be detected by a cross-correlation statistic $C(f)$ between two GW detectors~\cite{KAGRA:2021kbb,LIGOScientific:2019vic,LIGOScientific:2016jlg}.
The Gaussian likelihood for the measured cross-correlations $C(f)$ is 
\begin{equation}
\mathcal{L}\propto \exp{\left(-\frac{1}{2}\sum_k\frac{[C(f_k)-\Omega_{\rm{GW}}(f_{k};\bm{\theta})]^2}{\sigma^{2}(f_{k})}\right)},
\end{equation}
where the variance of $C(f)$ is
\begin{equation}
\sigma^{2}(f)\approx\frac{1}{2Tdf}\frac{P_{1}(f)P_{2}(f)}{\gamma_{T}^{2}(f)}\left(\frac{10\pi^{2}f^{3}}{3H_{0}^{2}}\right)^2.
\end{equation}
Here we assume one year's worth of data $T=1{\rm year}$ and a frequency bin width of $df=0.25[{\rm Hz}]$.
$P(f)_{1,2}$ are the one-sided noise power spectral densities of the two GW detectors, which are choosen as CE's~\cite{Curve:ce}.
Since the actual locations and arm orientations for CE are yet to be determined, we assume that CE shares the same overlap reduction function $\gamma_T(f)$ with LIGO.
As shown in Fig.~\ref{fig:OBBHs_fd}, the expected sensitivity of LIGO’s anticipated ``A+'' configuration~\cite{Thrane:2013oya,Ogw:code,Curve:A+} is not much lower than the total SGWB energy-density spectrum.
Although the sensitivity of CE is much lower than the total SGWB energy-density spectrum, there is no measured $C(f)$ for CE now.
Therefore, we just turn to the Fisher information matrix to obtain the forecasting constraints
\begin{equation}
\label{eq:F}
\mathcal{F}_{ij}=\sum_{k}\frac{1}{\sigma^{2}(f_{k})}\frac{\partial \Omega_{\rm{GW}}(f_{k};\bm{\theta})}{\partial \bm{\theta}_i}\frac{\partial\Omega_{\rm{GW}}(f_{k};\bm{\theta})}{\partial \bm{\theta}_j},
\end{equation}
where $\bm{\theta}$ is a vector consisting of $5$ free parameters
\begin{equation}
\nonumber
\bm{\theta}=\{\delta\omega_{220},\delta\tau_{220},z_{_R},z_{m_1},z_{q}\},
\end{equation}
and the derivative of the total SGWB energy-density spectrum with respect to each parameter is defined as
\begin{equation}\label{eqn-1} 
\frac{\partial\Omega_{\rm{GW}} }{\partial\bm{\theta}_{i}}=\frac{\Omega_{\rm{GW}}(\bm{\theta}_{i}+d\bm{\theta}_{i})-\Omega_{\rm{GW}}(\bm{\theta}_{i}-d\bm{\theta}_{i})}{2d\bm{\theta}_{i}},
\end{equation}
and is calculated numerically by choosing $d\bm{\theta}_{i}=0.1$.

\begin{figure*}[]
\begin{center}
\subfloat{\includegraphics[width=0.5\textwidth]{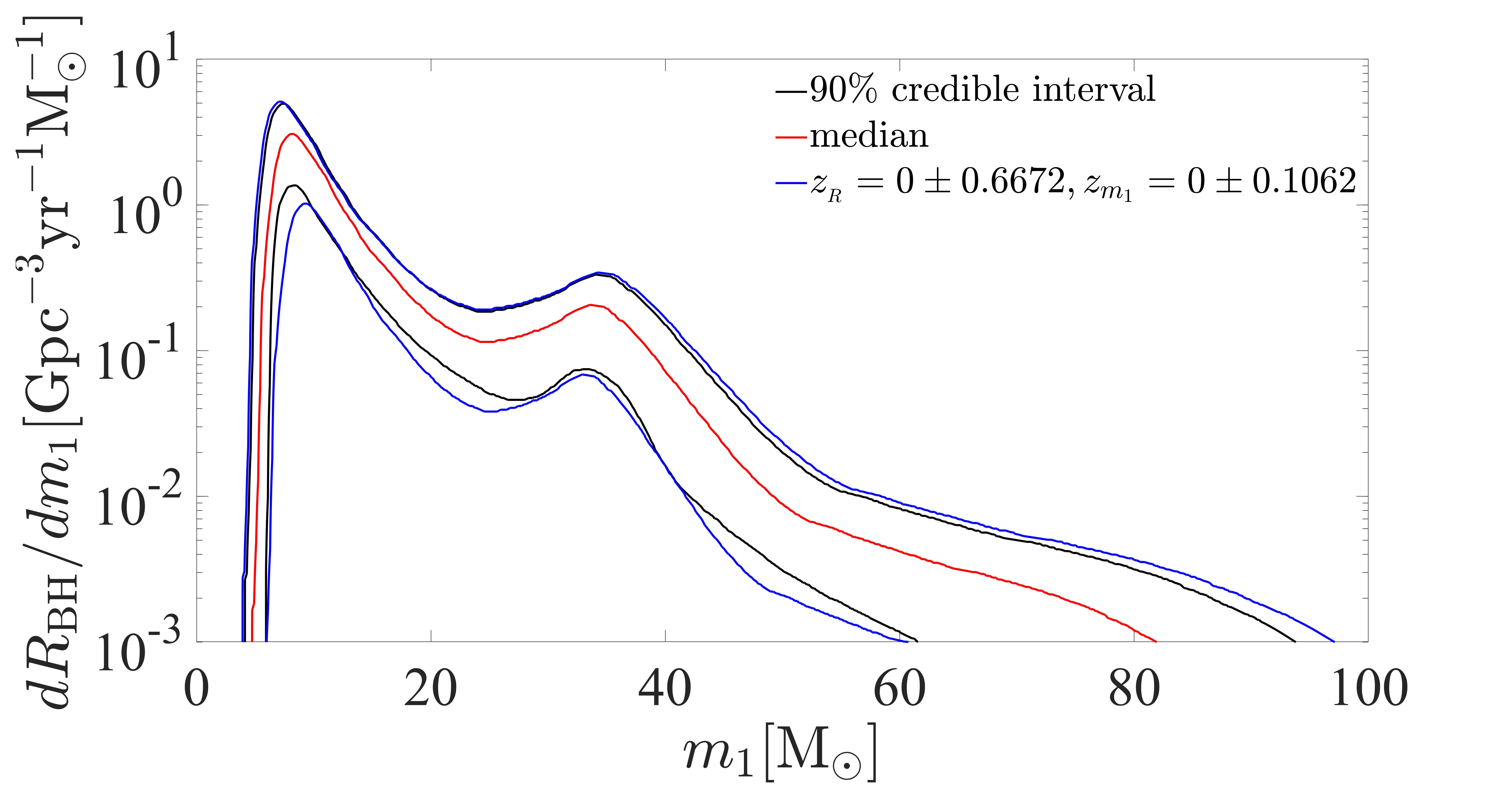}}
\subfloat{\includegraphics[width=0.5\textwidth]{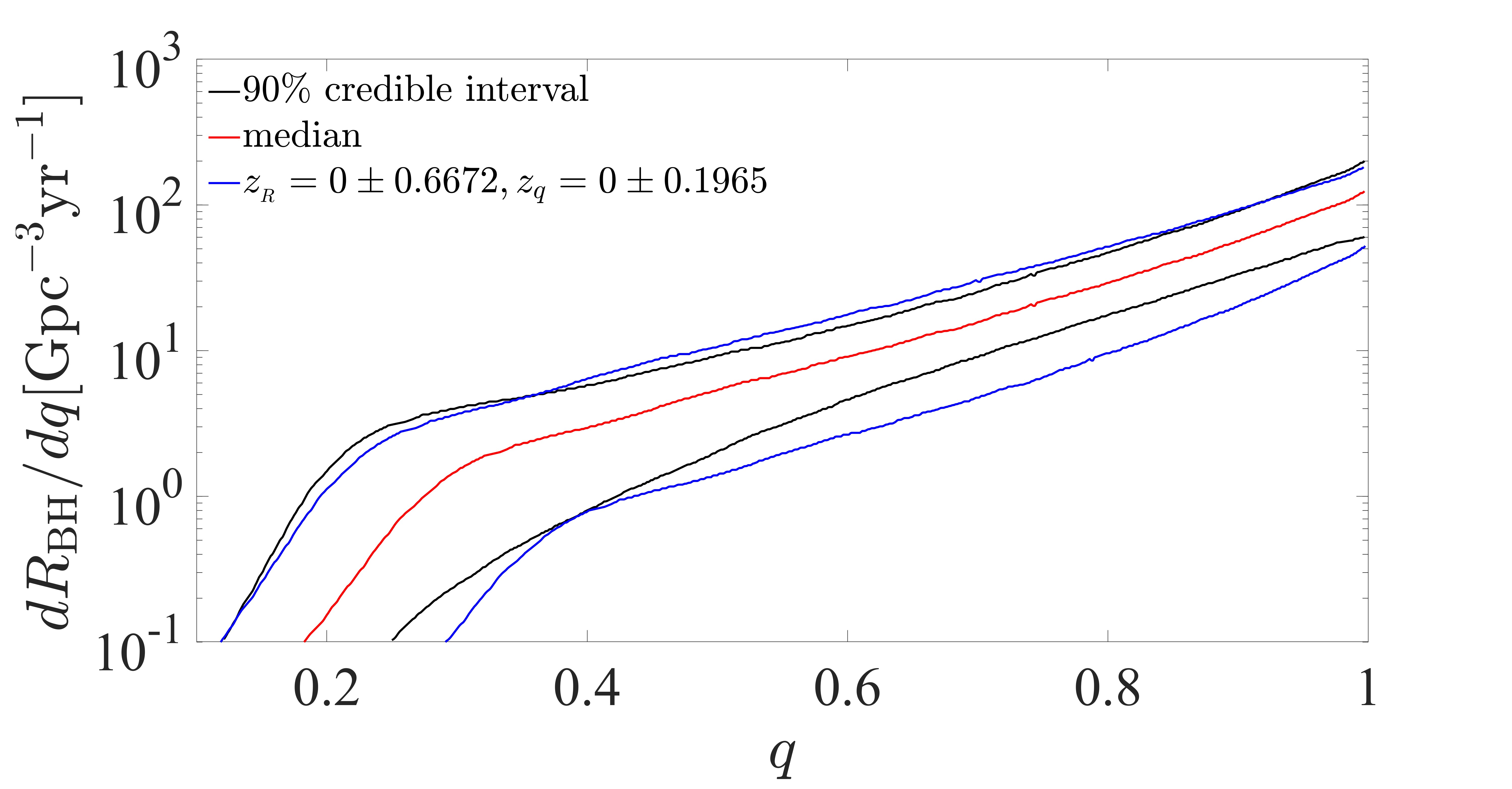}}
\end{center}
\captionsetup{justification=raggedright}
\caption{The primary mass (left) and mass ratio (right) distributions of the astrophysical BH binaries for the fiducial PP model at $z_{\rm quoted}=0.2$. The original constraints  from the fiducial PP model~\cite{KAGRA:2021duu} are the black curves. They are well mimicked by the blue ones which are obtained from the approximate median curves (red) by zooming in or out with parameters $\{z_{_R}=0\pm
0.6672,z_{m_{1}}=0\pm 0.1062,z_{q}=0\pm 0.1965\}$ ($90\%$ C.L.).}
\label{fig:zoom}
\end{figure*}

Here we introduce three new parameters $\{z_{_R},z_{m_1},z_{q}\}$ to summarize the original eight parameters of the fiducial PP model as
\begin{equation}
\label{eq:zoom} 
\begin{aligned}
&\frac{\mathrm{d}R_{\rm{BH}}}{\mathrm{d}m_{1}} \to A\frac{\mathrm{d}((1+z_{_R})R_{\rm{BH}})}{\mathrm{d}\left (m_{1}(\frac{m_{1}}{27.9[\rm{M_{\odot}}]})^{z_{m_{1}}}\right)},\\
&\frac{\mathrm{d}R_{\rm{BH}}}{\mathrm{d}q} \to B\frac{\mathrm{d} ((1+z_{_R})R_{\rm{BH}})}{\mathrm{d}\left (q(\frac{q}{1})^{z_{q}} \right )},\\
&A =\int \frac{\mathrm{d}R_{\rm{BH}}}{\mathrm{d}m_{1}} dm_{1}\left [\int \frac{\mathrm{d}R_{\rm{BH}}}{\mathrm{d}\left (m_{1}(\frac{m_{1}}{27.9[\rm{M_{\odot}}]})^{z_{m_{1}}}\right)} dm_{1}\right ]^{-1},\\
&B =\int \frac{\mathrm{d}R_{\rm{BH}}}{\mathrm{d}q} dq\left [\int \frac{\mathrm{d} R_{\rm{BH}}}{\mathrm{d}\left (q(\frac{q}{1})^{z_{q}} \right )}  dq\right ]^{-1},
\end{aligned}
\end{equation}
where $z_{_R}$ scales the merger rate $R_{\rm BH}(z_{\rm quoted}=0.2)$ directly, $z_{m_1}$ scales the primary mass distribution around the pivot mass $m_1=27.9[\rm{M_{\odot}}]$, $z_q$ scales the mass ratio distribution around the pivot ratio $q=1$ and $A$ (or $B$) is introduced to guarantee that $z_{m_1}$ (or $z_q$) doesn't modify $R_{\rm BH}(z_{\rm quoted}=0.2)$. That is to say, these three new parameters are independent of each other by construction. Therefore, we can obtain the non-flat priors of them from the constraints on the fiducial PP model as $\{z_{_R}=0\pm 0.6672,z_{m_{1}}=0\pm 0.1062,z_{q}=0\pm 0.1965\}$ ($90\%$ C.L.), where we have ignored the correlations between them. 
In Fig.~\ref{fig:zoom}, we show that the effects of $\{z_{_R},z_{m_{1}},z_{q}\}$ are similar to ones of the original fiducial PP model.

\begin{figure*}[]
\begin{center}
\subfloat{\includegraphics[width=0.5\textwidth]{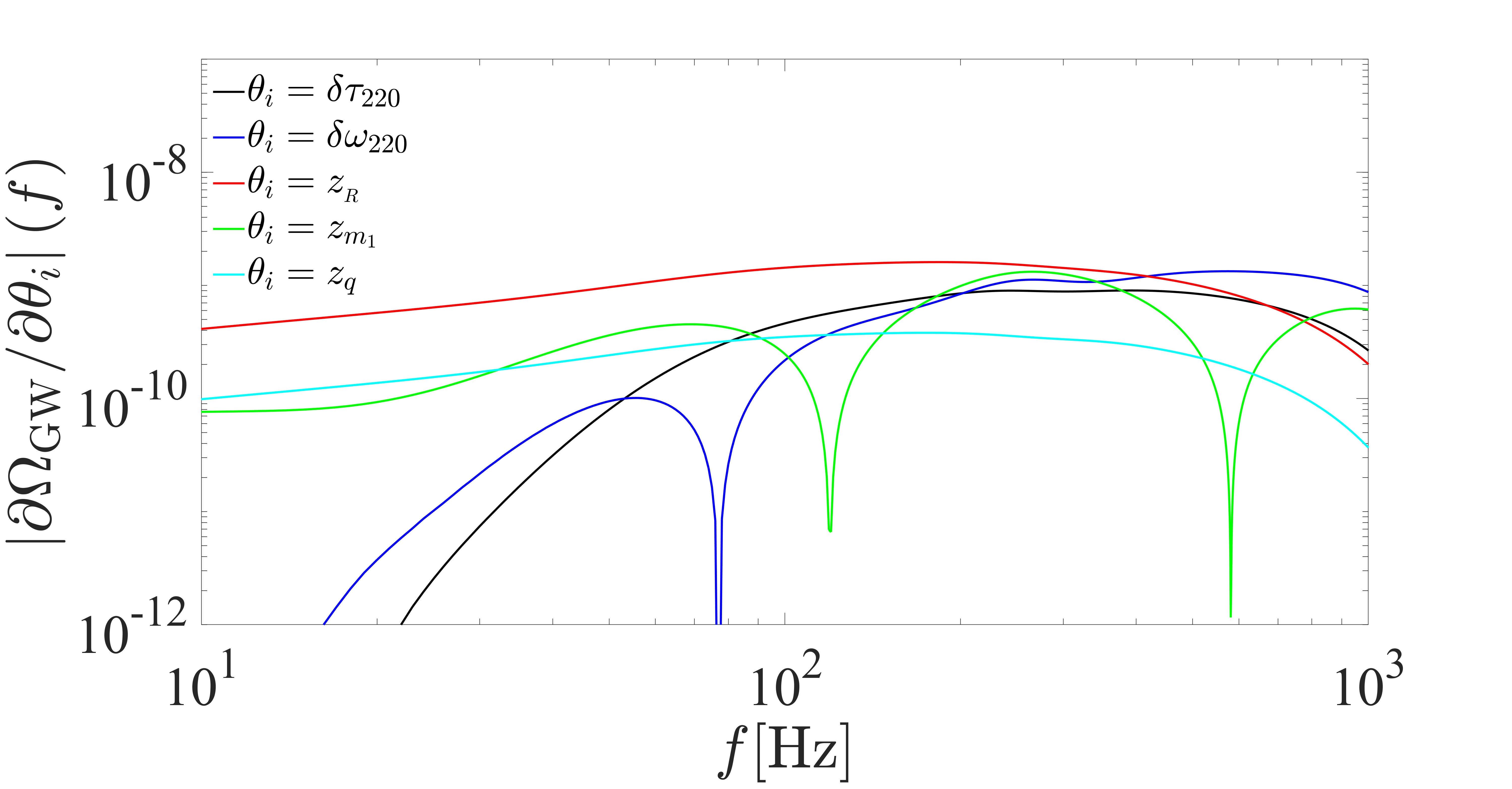}}
\end{center}
\captionsetup{justification=raggedright}
\caption{The absolute value of the derivatives of the total SGWB energy-density spectrum with respect to $\delta\tau_{220}$ (black curve), $\delta\omega_{220}$ (blue curve), $z_{_R}$ (red curve), $z_{m_{1}}$ (green curve) and $z_{q}$ (cyan curve) respectively.}
\label{fig:deriv}
\end{figure*}

We plot the absolute value of these derivatives in Fig.~\ref{fig:deriv}. While both $|\frac{\partial\Omega_{\rm{GW}}}{\partial\delta\tau_{220}}|$ and $|\frac{\partial\Omega_{\rm{GW}}}{\partial\delta\omega_{220}}|$ are suppressed with $f$ becoming lower which is consistent with the colored curves in Fig.~\ref{fig:OBBHs_fd}, $|\frac{\partial\Omega_{\rm{GW}}}{\partial z_{_R}}|$, $|\frac{\partial\Omega_{\rm{GW}}}{\partial z_{m_{1}}}|$ and $|\frac{\partial\Omega_{\rm{GW}}}{\partial z_{q}}|$ are not very sensitive to $f$. 
The root mean square errors of these parameters are given by
\begin{equation}
\sigma_i=\sqrt{(\mathcal{F}^{-1})_{ii}}.
\end{equation}
Since the observations of GW transients and the direct observation of SGWB are independent of each other, the constraints on $R_{\rm BH}$ from them should be independent of each other too. Here we rewrite the former constraints on $R_{\rm BH}$~\cite{KAGRA:2021duu} as a set of non-flat priors which can be added to $\mathcal{F}_{ij}$ in the form of
\begin{equation}
\begin{pmatrix}
 \sigma_{\delta \omega_{220}}^{-2}=0 &0  &0  &0  &0 \\
 0 &\sigma_{\delta \tau_{220}}^{-2}=0  &0  &0 &0 \\
 0 &0  &\sigma_{z_{_R}}^{-2}=6.1  &0  &0 \\
 0 &0  &0  &\sigma^{-2}_{z_{m_{1}}}=239.9  &0 \\
 0 &0  &0  &0  &\sigma_{z_{q}}^{-2}=70.1
\end{pmatrix}.
\end{equation}

\begin{figure*}[]
\begin{center}
\subfloat{\includegraphics[width=0.8\textwidth]{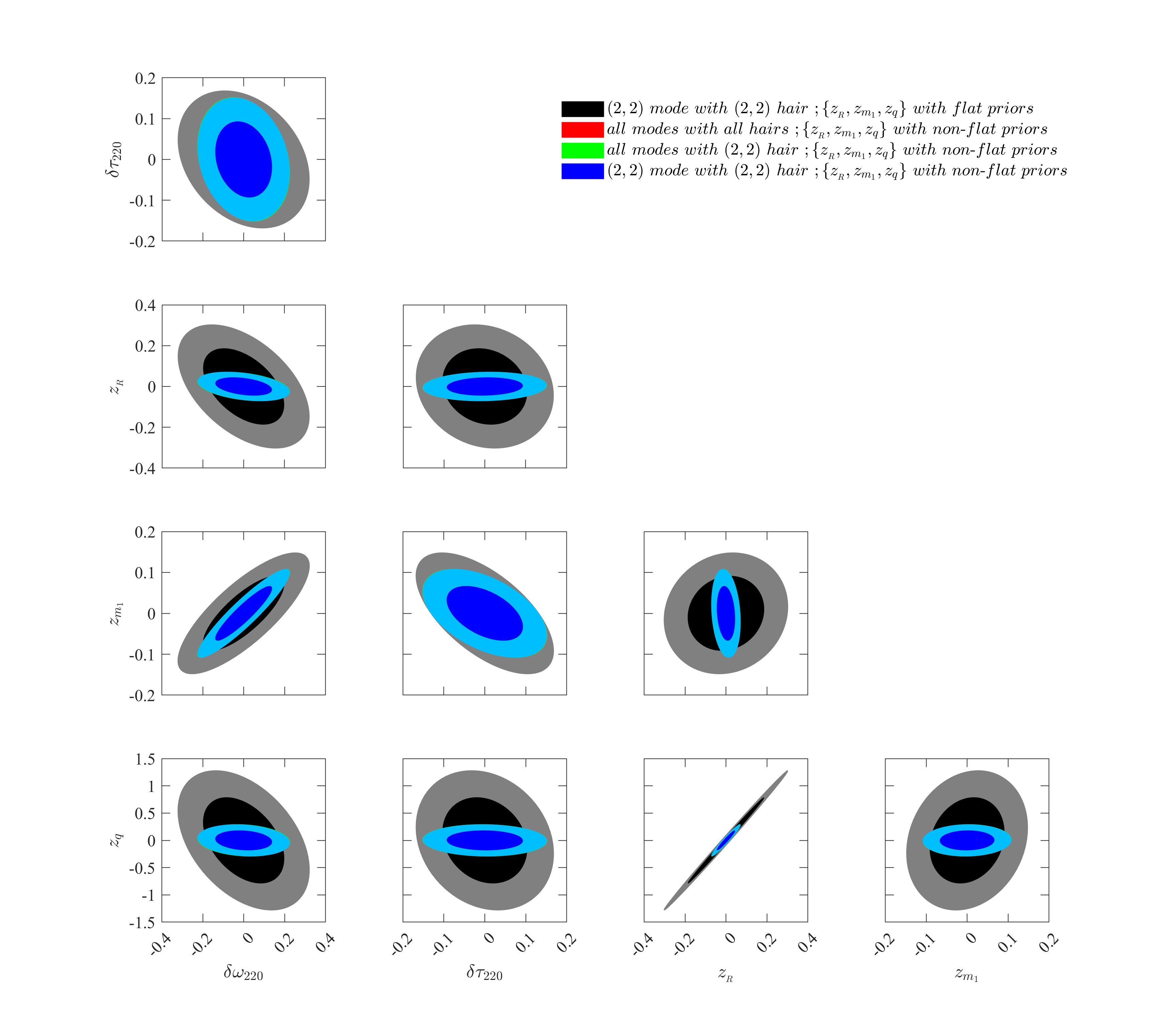}}
\end{center}
\captionsetup{justification=raggedright}
\caption{Error ellipses for $\{\delta\omega_{220},\delta\tau_{220},z_{_R},z_{m_{1}},z_{q}\}$, where $\delta\omega_{220}=0\pm0.1296$, $\delta\tau_{220}=0\pm0.0678$, $z_{_R}=0\pm 0.1219$, $z_{m_{1}}=0\pm 0.0597$ and $z_{q}=0\pm 0.5155$ at $68\%$ confidence range (inner black contour) when only $(2,2)$ mode and its hairs are considered and the flat priors of $\{z_{_R},z_{m_{1}},z_{q}\}$ are added, 
$\delta\omega_{220}=0\pm0.0903$, $\delta\tau_{220}=0\pm0.0608$, $z_{_R}=0\pm 0.0283$, $z_{m_{1}}=0\pm 0.0434$ and $z_{q}=0\pm 0.1161$ at $68\%$ confidence range (inner blue contour) when only $(2,2)$ mode and its hairs are considered and the non-flat priors of $\{z_{_R},z_{m_{1}},z_{q}\}$ are added, 
$\delta\omega_{220}=0\pm0.0907$, $\delta\tau_{220}=0\pm0.0610$, $z_{_R}=0\pm 0.0284$, $z_{m_{1}}=0\pm 0.0435$ and $z_{q}=0\pm 0.1167$ at $68\%$ confidence range (inner green contour) when the higher modes are also considered and the non-flat priors of $\{z_{_R},z_{m_{1}},z_{q}\}$ are added,
$\delta\omega_{220}=0\pm0.0908$, $\delta\tau_{220}=0\pm0.0611$, $z_{_R}=0\pm 0.0284$, $z_{m_{1}}=0\pm 0.0435$ and $z_{q}=0\pm 0.1167$ at $68\%$ confidence range (inner red contour) when the higher modes and their corresponding hairs are considered simultaneously and the non-flat priors of $\{z_{_R},z_{m_{1}},z_{q}\}$ are added 
and these outer lines are their corresponding $95\%$ confidence range.
It is worth noting that the red, green and blue contours almost completely overlap each other.} 
\label{fig:fisher}
\end{figure*}

In Fig.~\ref{fig:fisher}, we show the forecasting constraints on $\{\delta\omega_{220},\delta\tau_{220},z_{_R},z_{m_{1}},z_{q}\}$ with the $\mathtt{Fisher.py}$~\cite{Coe:2009xf} package, where $\delta\omega_{220}=0\pm0.1296$, $\delta\tau_{220}=0\pm0.0678$, $z_{_R}=0\pm 0.1219$, $z_{m_{1}}=0\pm 0.0597$ and $z_{q}=0\pm 0.5155$ at $68\%$ confidence range (inner black contour) when only $(2,2)$ mode and its hairs are considered and the flat priors of $\{z_{_R},z_{m_{1}},z_{q}\}$ are added,
$\delta\omega_{220}=0\pm0.0903$, $\delta\tau_{220}=0\pm0.0608$, $z_{_R}=0\pm 0.0283$, $z_{m_{1}}=0\pm 0.0434$ and $z_{q}=0\pm 0.1161$ at $68\%$ confidence range (inner blue contour) when only $(2,2)$ mode and its hairs are considered and the non-flat priors of $\{z_{_R},z_{m_{1}},z_{q}\}$ are added, 
$\delta\omega_{220}=0\pm0.0907$, $\delta\tau_{220}=0\pm0.0610$, $z_{_R}=0\pm 0.0284$, $z_{m_{1}}=0\pm 0.0435$ and $z_{q}=0\pm 0.1167$ at $68\%$ confidence range (inner green contour) when the higher modes are also considered according to Eq.~(\ref{eq:hm}) and the non-flat priors of $\{z_{_R},z_{m_{1}},z_{q}\}$ are added
and $\delta\omega_{220}=0\pm0.0908$, $\delta\tau_{220}=0\pm0.0611$, $z_{_R}=0\pm 0.0284$, $z_{m_{1}}=0\pm 0.0435$ and $z_{q}=0\pm 0.1167$ at $68\%$ confidence range (inner red contour) when the higher modes and their corresponding hairs are considered simultaneously and the non-flat priors of $\{z_{_R},z_{m_{1}},z_{q}\}$ are added. 

We can find that the green contours almost completely overlap the blue contours because the higher modes without hairs just shift the values of $\Omega_{\rm GW}(f)$ but don't affect their derivative with respect to each hair $\frac{\partial\Omega_{\rm{GW}} }{\partial\bm{\theta}_{i}}(f)$. Since the contributions of the higher modes' hairs should be smaller than the contributions of the higher modes themselves, the red contours also almost completely overlap the blue contours.
Interestingly, there is a slight correlation between $(2,2)$ mode's two hairs. It means that certain deviations from GR probably affect the ringdown frequencies and damping times simultaneously.
There is an obvious positive correlation between $z_{_R}$ and $z_{q}$. Although $(1+z_{_R})$ serves as a factor of $R_{\rm{BH}}$ (or $\Omega_{\rm{GW,BH}}$) according to Eq.~(\ref{eq:zoom}) (or Eq.~(\ref{eq:o_k})), $z_{q}$ doesn't affect $R_{\rm{BH}}$ directly according to the definition of $B$ in Eq.~(\ref{eq:zoom}). Therefore, $z_{q}$ must affects $\Omega_{\rm{GW,BH}}$ via $\left \langle dE_{s}/df_{s}\right \rangle_{\rm{BH}}$ according to Eq.~(\ref{eq:mdf}). For example, $z_{q}=0.1965$ widens the mass ratio distribution to $0.1 \lesssim q\le 1$ as shown in the right subplot of Fig.~\ref{fig:zoom} then suppresses the probability of $0.3\lesssim q\le 1$ overall due to $B$, hence a smaller $\left \langle dE_{s}/df_{s}\right \rangle_{\rm{BH}}$ due to the suppression of $dE_{s}/df_{s}$ by the smaller mass ratio $0.1 \lesssim q \lesssim 0.3$. For $z_{q}=-0.1965$, the reverse applies.
There are correlations between $\{z_{_R},z_{q}\}$ and $\{\delta\omega_{220},\delta\tau_{220}\}$ in the black contours when the flat priors of $\{z_{_R},z_{q}\}$ are added but these correlations almost disappear in the blue contours when the non-flat priors of $\{z_{_R},z_{q}\}$ are added. 
The reason for this disappearance is that both $z_{_R}$ and $z_{q}$ can serve as a factor of $\Omega_{\rm{GW,BH}}$ and make an almost frequency-independent contribution to $\Omega_{\rm{GW,BH}}$.
That is to say, the absolute contribution of $\{ z_{_R}=0\pm 0.0283, z_{q}=0\pm 0.1161\}$ is much larger at lower frequency but much smaller at higher frequency than that of $\{\delta\omega_{220}=0\pm0.0903, \delta\tau_{220}=0\pm0.0608\}$, as hinted in Fig.~\ref{fig:deriv}.
These totally different behaviors of them lead to the negligible correlations between them.
%The reason for this disappearance is that both $z_{_R}$ and $z_{q}$ can serve as a factor of $\Omega_{\rm{GW,BH}}$ whose derivatives with respect to $(2,2)$ mode's two hairs are hardly affected by tiny values of $z_{_R}$ and $z_{q}$.
There are obvious correlations between $z_{m_{1}}$ and $\{\delta\omega_{220},\delta\tau_{220}\}$ in the black and blue contours. It means that $\Omega_{\rm{GW,BH}}$ at higher frequency is sensitive to $z_{m_{1}}$ which results from the dramatic influence of $z_{m_{1}}$ on the primary mass distribution around $m_{1}\lesssim10[\rm{M_{\odot}}] $, hence indirect correlations between $z_{m_{1}}$ and $\{\delta\omega_{220},\delta\tau_{220}\}$ at higher frequency. 
Finally, the non-flat prior of $z_{m_{1}}$ from the present observations of GW transients is similar to the its forecasting constraint from the future observation of SGWB only, which leads to two similar constraints in the $z_{m_{1}}-\delta\omega_{220}$ plane and $z_{m_{1}}-\delta\tau_{220}$ plane respectively.

\section{Summary and Discussion}
\label{sec:SD}
In this paper, we assume that the effects of the extra hairs on the inspiral and merger regions can be summarized by a set of effective intrinsic parameters under the no-hair theorem and then the residual effects of these extra hairs will just appear during an effective ringdown region. Then we turn to the $\mathtt{PyCBC}$ package\cite{Biwer:2018osg} and the $\mathtt{pSEOBNRv4HM\_PA}$ model~\cite{Ghosh:2021mrv,Mihaylov:2021bpf} to obtain the effective complete IMR time-domain waveform with hairs, as shown in Fig.~\ref{fig:gw150914_td}. 
After the Fourier transform (as shown in Fig.~\ref{fig:gw150914_fd}), we calculate the energy spectrum for every possible BH binary with different primary mass and mass ratio individually, as shown in Fig.~\ref{fig:dedf150914_fd}.
Combining these energy spectra under a fixed fiducial PP model~\cite{LIGOScientific:2020kqk,KAGRA:2021duu} , we obtain the modified total SGWB energy-density spectrum at frequency $10[{\rm Hz}]\lesssim f\lesssim10^3[{\rm Hz}]$ for given hairs, as shown in Fig.~\ref{fig:OBBHs_fd}. Here we suppose that the ratio of the contribution of BH binary mergers to the contribution of NS binary and NS–BH mergers is about $1:0.3$~\cite{KAGRA:2021duu}. 
To further take the uncertainties of the fiducial PP model~\cite{LIGOScientific:2020kqk,KAGRA:2021duu} into consideration, the Fisher information matrix should also include the parameters of the fiducial PP model. For simplicity, we reduce the original eight parameters of the fiducial PP model to $\{z_{_R},z_{m_{1}},z_{q}\}$ and change the uncertainties of the original eight ones to the non-flat priors of the latter three ones.
By choosing the all free parameters as $\sim0.1$, we calculate numerically the derivative of the total SGWB energy-density spectrum with respect to each one respectively, as shown in Fig.~\ref{fig:deriv}.
To obtain the variance of SGWB, we assume that CE shares the same overlap reduction function with LIGO.
Finally, the forecasting constraints on hairs at $68\%$ confidence range are $\delta\omega_{220}=0\pm0.1296$ and $\delta\tau_{220}=0\pm0.0678$ when the flat priors of $\{z_{_R},z_{m_{1}},z_{q}\}$ are added but $\delta\omega_{220}=0\pm0.0903$ and $\delta\tau_{220}=0\pm0.0608$ when the non-flat priors of $\{z_{_R},z_{m_{1}},z_{q}\}$ are added, as shown in Fig.~\ref{fig:fisher}. As for the higher modes, we find that they hardly affect the forecasting constraints on $\{\delta\omega_{220},\delta\tau_{220}\}$ while they do contribute to and shift the total SGWB energy-density spectrum. And so do their corresponding hairs.

There are three caveats. The first one is that we turn to the $\mathtt{pSEOBNRv4HM\_PA}$ model~\cite{Ghosh:2021mrv,Mihaylov:2021bpf} where the extra hairs appear only during the ringdown region. In fact, we have mimicked the effects of the extra hairs on the inspiral and merger regions with an effective GR's IMR waveform and then leaved the residual effects on the ringdown region alone. Therefore, the chosen fiducial PP model~\cite{LIGOScientific:2020kqk,KAGRA:2021duu} for
the mass distribution is also an effective one which has included the effects of the extra hairs on the inspiral and merger regions. The second one is that we assume the ratio of the contribution of BH binary mergers to the contribution of NS binary and NS–BH mergers is about $1:0.3$ according to Fig.~23 of \cite{KAGRA:2021duu}. It is just a
temporary assumption and will be improved according to the newest GW observations. The third one is that CE shares the same overlap reduction function $\gamma_T(f)$ with LIGO. It is also a temporary assumption. Because $\gamma_T(f)$ is determined by the relative positions and orientations of a pair of detectors and the actual locations and arm orientations for CE are yet to be determined. 

\begin{acknowledgments}
Ke Wang is supported by grants from NSFC (grant No. 12005084 and grant No.12247101).
\end{acknowledgments}

\end{document}